\documentclass[conference]{IEEEtran}
\IEEEoverridecommandlockouts
\usepackage{cite}
\usepackage{amsmath,amssymb,amsfonts}
\usepackage{algorithmic}
\usepackage{graphicx}
\usepackage{textcomp}
\usepackage{xcolor}
\usepackage{url}
\usepackage{amsmath}
\usepackage{array}
\usepackage{booktabs}
\usepackage{fullpage}
\usepackage{graphicx}
\usepackage{threeparttable}
\usepackage{wasysym}

\def\BibTeX{{\rm B\kern-.05em{\sc i\kern-.025em b}\kern-.08em
    T\kern-.1667em\lower.7ex\hbox{E}\kern-.125emX}}
\begin{document}

\title{vFHE: Verifiable Fully Homomorphic Encryption with Blind Hash\\
}

\author{\textsuperscript{1}Qian Lou, \textsuperscript{2}Muhammad Santriaji, \textsuperscript{2}Ardhi Wiratama Baskara Yudha, \textsuperscript{2}Jiaqi Xue, and \textsuperscript{1}Yan Solihin\\
University of Central Florida\\
\textsuperscript{1}\{qian.lou, yan.solihin\}@ucf.edu,
\textsuperscript{2}\{santriaji, yudha, jiaq\_xue\}@knights.ucf.edu
}

\maketitle

\begin{abstract}
Fully homomorphic encryption (FHE) is a powerful encryption technique that allows for computation to be performed on ciphertext without the need for decryption. FHE will thus enable privacy-preserving computation and a wide range of applications,  such as secure cloud computing on sensitive medical and financial data, secure machine learning, etc. 
Prior research in FHE has largely concentrated on improving its speed, and great stride has been made. However, there has been a scarcity of research on addressing a major challenge of FHE computation: client-side data owners cannot verify the integrity of the calculations performed by the service and computation providers, hence cannot be assured of the correctness of computation results. This is particularly concerning when the service or computation provider may act  in an untrustworthy, unreliable, or malicious manner and tampers the computational results. Prior work on ensuring FHE computational integrity has been {\em non-universal} or {\em incurring too much overhead}. We propose vFHE to add computational integrity to FHE without losing universality and without incurring high performance overheads. 
\end{abstract}

\begin{IEEEkeywords}
Fully Homomorphic Encryption; Integrity; Blind Hash; Secure Computation;
\end{IEEEkeywords}

\section{Introduction}
Privacy-preserving technology in cloud computing is crucial, as it allows for deploying cloud-based applications where data privacy is paramount. This property is especially important in cases where the data is highly sensitive or when compliance with increasingly stringent privacy regulations is required. Fully Homomorphic Encryption (FHE)~\cite{gentry2009fully, brakerski2014leveled, halevi2019BFVimproved, kim2022approximate} is a distinct privacy-preserving technique that enables computation on encrypted data without the need for decryption. 

FHE enables privacy-preserving computation in a variety of applications. For example, data owners such as Alice can gain new insights from their private data through service providers such as Bob, who can perform computations, manipulations, and even aggregations on the data without access. This technique has numerous use cases, such as secure cloud computing for sensitive medical and financial data~\cite{AmazonScience, IBMSecurity, GoogleCloud, Google, Microsoft,reagen2021cheetah} and secure machine learning~\cite{graepel2013ml,lu2016using, chen2018logistic,gilad2016cryptonets, brutzkus2019low}.

\begin{figure}[t!]
\centerline{
  \hspace{0.3cm}
\includegraphics[width=0.5\textwidth]{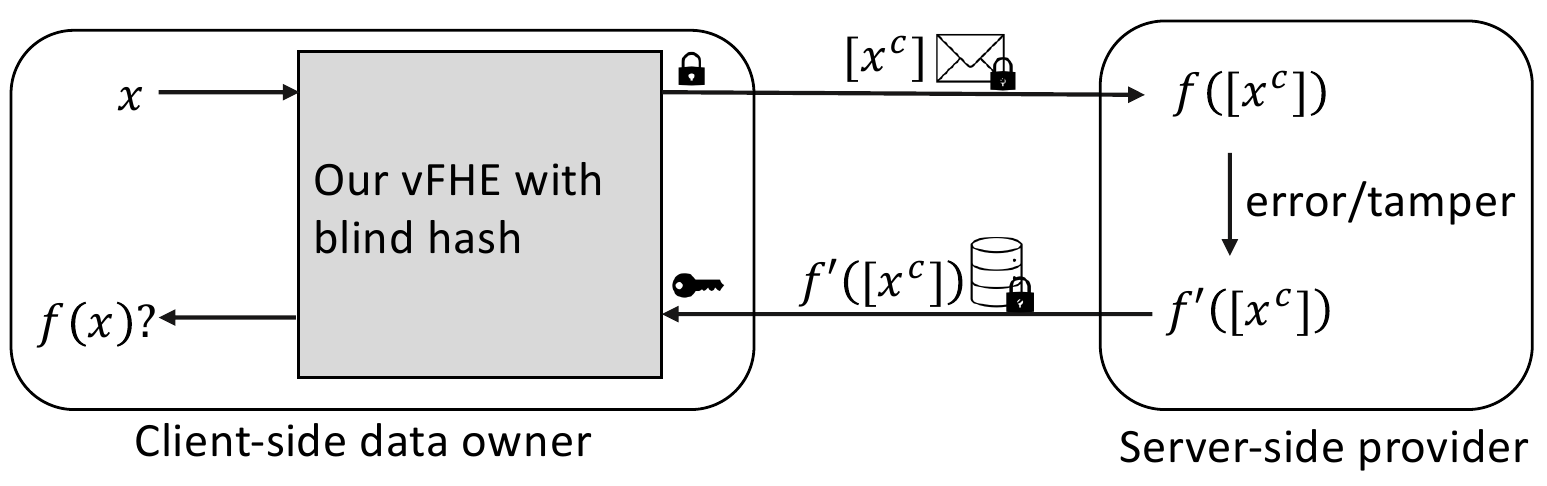}}
\caption{Illustrating vFHE with blind hash, which enables the verification of Fully Homomorphic Encryption (FHE) against both malicious tampering and computational errors. }
\label{fig:overview}
\end{figure}

Prior research efforts in FHE have mainly concentrated on boosting its speed, leading to notable advancements in this domain. For instance, recent studies~\cite{CryptoNets:ICML2016, lou2019she, lou2020safenet, lou-2021-cryptogru, lou2020autoprivacy, lou2020falcon, Brutzkus:ICML19, GAZELLE:USENIX18, Pratyush:USENIX2020} have improved the efficiency of FHE through innovative schemes and hardware acceleration, such as the use of GPU support for FHE~\cite{fhe-gpu} and ciphertext batching~\cite{GAZELLE:USENIX18, Pratyush:USENIX2020}, which substantially reduce latency and enhance throughput for privacy-preserving computation.

Despite the significant progress in FHE research, there has been a scarcity of attention given to one of its major challenges: the inability of client-side data owners to verify the integrity of computations performed by service and computation providers, leading to uncertainties about the accuracy of results. This concern is especially pressing in cases where the provider is untrustworthy, unreliable, or acts maliciously and may tamper with the computational results.

\newcommand*\feature[1]{\ifcase#1 -\or\LEFTcircle\or\CIRCLE\fi}
\newcommand*\f[1]{\feature#1}
\makeatletter
\newcommand*\ex[7]{#1&\f#2&\f#3&\f#4&\f#5&\f#6&\f#7}
\begin{table*}[h]
\centering
\footnotesize
\begin{threeparttable}
\caption{Comparison of current FHE integrity schemes and our Blind Hash method}
\label{tab:features}
\setlength{\tabcolsep}{5pt}
\begin{tabular}{lcccccc}
\toprule
\textbf{Scheme} & \textbf{Universality} & \textbf{Scalability} & \textbf{Security} & \textbf{Low Overhead} & \textbf{Verifiable Model}  & \textbf{No hardware}  \\
\midrule
%\midrule
%\ex{MAC}{1}{0}{2}{0}{0}{2}\\
\ex{MAC'~\cite{chatel2022verifiable}} {1}{0}{2}{0}{0}{2}\\
\ex{Residue\cite{Awadallah2021}  \cite{cryptoeprint:2023/231} }{2}{0}{0}{0}{0}{2}\\
%\ex{ZKP}{1}{0}{2}{0}{2}{2}\\
\ex{ZKP' ~\cite{viand2023verifiable}}{1}{0}{2}{0}{2}{2}\\
\ex{TEE~\cite{natarajan2021chex, viand2023verifiable} }{2}{0}{2}{0}{2}{0}\\
\ex{\textbf{Blind Hash}}{2}{2}{2}{2}{2}{2}\\
\bottomrule
\end{tabular}
\begin{tablenotes}

\begin{small}
\item $\feature2=\text{provides property}$; $\feature1=\text{partially provides property}$;
$\text{\feature0}=\text{does not provide property}$;
%\item \textsuperscript{\dag}has academic publication;
%\textsuperscript{}end-user tool available
\end{small}
\end{tablenotes}
\end{threeparttable}
\end{table*}

Prior work ensuring FHE computational integrity has been {\em non-universal} or {\em incurring too much overhead}. The majority of research focuses on two approaches: cryptographic integrity checking protocol~\cite{bhadauria2020ligero++, bunz2018bulletproofs, gennaro2010non,goldwasser2015delegating, parno2016pinocchio, weng2021wolverine, brakerski2011fully, bois2021flexible, fiore2014efficiently, fiore2020boosting, ganesh2021rinocchio}, including homomorphic message authentication code (MAC and MAC'\cite{chatel2022verifiable}) and zero-knowledge proofs (ZKP and ZKP'\cite{viand2023verifiable}); and relying on trusted execution environment (TEE) hardware~\cite{natarajan2021chex, wang2019toward, coppolino2020vise,sabt2015trusted}. The former approach suffers from non-universality from incompatibility with ring-based FHE schemes. Different FHE schemes, such as BGV~\cite{brakerski2014leveled} and CKKS~\cite{fhe-ckks}, operate over different encoding methods, polynomial structures, and ciphertext space. Thus a cryptographic integrity method, e.g., MAC, MAC', ZKP', that alters the inner encoding and algorithms in a specific FHE scheme needs redesign for a different scheme. The generic ZKP without optimization for a specific FHE scheme suffers from a large overhead. Modifying cryptographic integrity to make it efficient and compatible with all modern FHE schemes without impacting FHE functionalities has so far remained elusive~\cite{chatel2022verifiable}.

% \aji{Prior work on ensuring FHE computational integrity has been incuring too much overhead. The majority of research is focusing on three approaches: Message Authentication Code (MAC), Zero Knowledge Proof (ZKP) and relying on Trusted Execution Environment (TEE). MAC approach suffers large overhead because it requires an interactive authentication to bypass non linear function. Meanwhile, ZKP would suffer when it is used for evaluating highly configurable function. The keys (sk,pk) of ZKP is bind only for one function configuration and needs to be regenerated for different configuration. While deploying a FHE to TEE will ensure integrity, it has a high overhead because of double encryption.}

%A generic and efficient integrity verification that is compatible with various FHE schemes is needed. We propose a \textit{blind hash} method to enable verifiable FHE as Figure~\ref{fig:overview} shows.
For these reasons, there is a need for a generic and efficient integrity verification solution that is compatible with various FHE schemes. In this context, we propose a novel approach called the \textit{blind hash} method, which enables verifiable FHE. The proposed method is based on a plug-in algorithm that can be used with different FHE schemes. The method involves generating a blind hash of the raw data, which is then used to verify the integrity of the computation. We present our vFHE module with blind hash  in Figure~\ref{fig:overview}. In the context of a client-side data owner and a server-side service/computation provider, let's assume that the data owner wants to compute a function $f()$ over an input $x$ resulting in $f(x)$. In order to do so, the raw data $x$ is preprocessed by our proposed \textit{blind hash} function to generate $x_c$. This $x_c$ can be encoded and encrypted into ciphertext $[x_c]$ using any fully homomorphic encryption (FHE) schemes. The server can perform the function $f([x_c])$ without decrypting the data. The encrypted results obtained from the server side can be decrypted by the data owner. Our proposed verifiable FHE (vFHE) method allows for verification of the integrity and correctness of $f(x)$ against any computational errors or malicious tampering. Specifically, if the server-side computation results in $f'([x_c]) = f([x_c])$ where $f'()$ is a tampered result that yields a different output than $f()$, the vFHE method would detect this and alert the data owner of the tampering attempt. 

% \aji{Table I. I remove implementation because all of latest paper has implementation already, I am not sure about Universality, latest paper on FHE-ZKP already provide universality, I remove ML support because blind hash (excluding blind rotation) doesn't support ML}

%\section{Motivational Example}
%\input{motivational.tex}

\section{Background and Related Work}
In this section, we briefly describe FHE and the current state of the art to guarantee integrity in FHE.
\subsection{Fully Homomorphic Encryption}
An encryption scheme transforms plaintext (unencrypted data) into ciphertext (encrypted data) using an encryption algorithm to make the ciphertext unintelligible to unauthorized parties. Encryption schemes are used to protect the confidentiality and privacy of data, as well as to ensure the integrity and authenticity of data. FHE stands for Fully Homomorphic Encryption. This type of encryption scheme allows computations to be performed on ciphertexts without first decrypting them. In other words, it enables the computation of functions directly on encrypted data.

Currently, various Fully HE (FHE) schemes such as BGV~\cite{brakerski2014leveled} and BFV~\cite{halevi2019BFVimproved}, as well as CKKS~\cite{fhe-ckks}, are based on ring-based learning with errors and operate on different polynomials in a ring-based structure. Since FHE operations are approximately 10,000 times slower than non-FHE operations~\cite{CryptoNets:ICML2016}, a significant portion of FHE and FHE-based privacy-preserving machine learning research~\cite{CryptoNets:ICML2016, lou2019she,lou2020safenet,Brutzkus:ICML19, GAZELLE:USENIX18, Pratyush:USENIX2020} has focused on improving the efficiency of FHE through innovative schemes and hardware acceleration, such as incorporating GPU support for FHE~\cite{fhe-gpu} and utilizing techniques like ciphertext batching~\cite{GAZELLE:USENIX18, Pratyush:USENIX2020} which significantly reduce latency and improve the throughput of privacy-preserving computation.

\subsection{Comparison with Related Works}

In table \ref{tab:features}, we compare our blind hash method with the related works.  Research on computational integrity primarily focuses on two approaches: cryptographic integrity checking protocols~\cite{bhadauria2020ligero++, bunz2018bulletproofs, gennaro2010non,goldwasser2015delegating, parno2016pinocchio, weng2021wolverine, brakerski2011fully, bois2021flexible, fiore2014efficiently, fiore2020boosting, ganesh2021rinocchio}, e.g., MAC, MAC', ZKP, and ZKP'; and utilizing trusted execution environments (TEE) hardware~\cite{natarajan2021chex, wang2019toward, coppolino2020vise,sabt2015trusted}. 

Several studies have presented MAC-based approaches in scholarly literature~\cite{brakerski2011fully, bois2021flexible, fiore2014efficiently, fiore2020boosting, ganesh2021rinocchio}. A notable example of a MAC~\cite{MAC-CCS2014} focuses on verifying ciphertext computations by specifically examining quadratic functions within a particular variant of the BV scheme~\cite{brakerski2011fully}. However, these methods are not universally applicable to all FHE schemes and have limitations in scalability, efficiency and functionality~\cite{chatel2022verifiable}.
In comparison, the MAC' method~\cite{chatel2022verifiable} improves efficiency and universality by implementing alternative encoding methods. Nonetheless, the applicability of this method to all FHE schemes universally remains uncertain, in contrast to our vFHE approach, which does not necessitate alterations to encoding techniques. Additionally, the overhead associated with this method is no less than twice that of the original computation.

Modulus residue approaches \cite{Awadallah2021}  \cite{cryptoeprint:2023/231} are derivations of the MAC scheme. They offer lower overhead by creating an unencrypted verification encoding with reduced data size. The efficiency of these residue methods, which feature unencrypted verification and an unprotected verification function, is achieved at the expense of compromised security. Common FHE schemes, such as BFV, BGV, and CKKS, enable FHE computation with plaintext. A malicious actor could manipulate the data and bypass the verification by simply altering both the data and the checksum through arithmetic plaintext operations on encrypted data. In comparison, our blind hash scheme maintains security while minimizing overhead. Although these residue methods are well-suited for safeguarding integrity against faulty hardware, they are not as effective against sophisticated adversaries. Despite their significantly lower overhead relative to MAC schemes, their limited scalability arises from the linear increase in verification size as the message size grows.

The ZKP method \cite{ganesh2021rinocchio} has been explored for maintaining computational integrity; however, employing a Zero-Knowledge Proof technique for universal FHE verification without optimization specific to a particular FHE scheme results in considerable overhead. This overhead intensifies in proportion to the size of the data. In contrast, ZKP' \cite{viand2023verifiable}, optimized for a distinct FHE scheme, might not attain full universality.

The hardware-based strategy, which executes complete or partial FHE operations within a Trusted Execution Environment (TEE) \cite{natarajan2021chex, viand2023verifiable}, exhibits universality; however, it imposes considerable performance overheads. The reported overheads range from $3-30\times$ \cite{natarajan2021chex, wang2019toward, coppolino2020vise, chatel2022verifiable}, highlighting the significant performance implications of this approach. This is due to one key limitation: TEEs were designed for non-FHE computation. Hence their designs mismatch with what FHE needs. %This calls for a {\em clean-slate redesign} of TEEs for supporting FHE computation to reduce the performance overheads.

\section{Threat Model}
We consider an outsourcing scheme between a client-side data owner $\mathcal{C}$ and a server $\mathcal{S}$, where $\mathcal{S}$ executes a function $f(x): X \rightarrow Y$ on data provided by $\mathcal{C}$. The function $f$ can either belong to $\mathcal{C}$ (e.g., in IaaS/PaaS~\cite{Armbrust2010}), or to the server (e.g., in SaaS/API~\cite{Armbrust2010}). We adopt a more realistic threat model in which the server $\mathcal{S}$ is not fully trustworthy and may be malicious or vulnerable to tampering with the ciphertext $f(x)$. This departs from the traditional semi-honest setting, in which $\mathcal{S}$ is assumed to be honest but curious (HBC) about inferring $\mathcal{C}$'s data privacy. Our scheme should  satisfy the following security properties: {\bf Integrity}: $\mathcal{C}$ could detect an integrity attack when interacting with $\mathcal{S}$ for any input $x$ and ensure the correctness of $y=f(x)$.  {\bf Data Privacy}: $\mathcal{S}$ cannot learn any information about input $x$. 
While the function $f()$ can be made public to both the client and server in many applications~\cite{Occlumency:TEE-deep-learning,VISE-TEE-FHE}, it is still important to note that in certain applications and scenarios~\cite{tramer2018slalom}, preserving the privacy of the function may be a critical concern. {\bf Function Privacy}: If $f()$ is provided by $\mathcal{C}$, $\mathcal{S}$ cannot learn information about $f()$ beyond its size. If $f()$ belongs to $\mathcal{S}$, $\mathcal{C}$ cannot learn more about $f()$ than what is revealed by $y = f(x)$.

\section{vFHE Design}

\textbf{Design Principle.} Our {\em blind hash} scheme was initially inspired by  algorithm-based fault tolerance (ABFT) techniques~\cite{hari2021making-checksum, ABFT-1984-checksum, ding2011matrix-2011-checksum, checksum-2017} that utilize checksums to detect computational errors due to faults in systems. The checksum provides redundant mathematical relationship of data that is preserved in the output, such that computational error becomes detectable through verifying the relationship in the output. However, while effective in detecting faults, ABFT techniques have a fundamental flaw if used for detecting errors due to attacks in untrusted environments: attackers can bypass the detection by manipulating both data and the checksum, leveraging the knowledge of the computation process. Even though the checksum is encrypted by FHE, it is mathematically related to data in ways known or guessable by the attacker, creating a security gap between fault detection and attack detection. Our {\em blind hash} bridges the gap and improves upon ABFT by following the main principle of {\em concealing the checksum computing process} from the untrusted environments: {\em blind hash} incorporates an extra layer of security by adding a blind hash function into the checksum computation process. The blind hash function is only visible to the data owners and generated checksum is encrypted by FHE, making it secure and tamper-resistant. {\em Blind hash} not only detects faults, as in ABFT, but also detects attacks. 

\textbf{Workflow.} Our development of {\em blind hash} will take the following steps: (i) blind hash calculation. Given data $x$ and predefined hash vector $h^x$, the data owner calculates the blind hash checksum values by $hash(x, h^x)=h^xx$, attaches it to the original data as $x'$, and encrypts $x'$ using FHE to generate $[x']$, here $[x']$ represents ciphertext of plaintext $x'$.   (ii) data sharing. The data owner shares encrypted $[x']$ with the service provider. (iii) private computation for result and proof. The service provider performs arithmetic function $[f([x'])]$ on encrypted data $[x']$ directly, where $f()$ function can be any arithmetic function, including convolution, matrix operation, etc, and it usually depends on parameters $w$, e.g., convolution filter, machine learning weights.  

(iv) client's integrity checking. The client decrypts $[f([x'])]$, i.e., ($[f([hash(x, h^x)])]$ and $[f([x])]$), and verifies the integrity and correctness by checking the results and proof $f([hash(x, h^x)])$ and $hash(h^x, f([x]))$.

\textbf{Function Privacy.} The function $[f([x'], w)]$ can be directly performed in many applications~\cite{Occlumency:TEE-deep-learning,VISE-TEE-FHE} where the variable $w$ is public to both the client and the server. However, for applications where the weight $w$ is kept confidential by the client, the client can compute the blind hash on $w$ and encrypt it in the same manner as the input $x$. The encrypted $w'$ can be shared with the server, allowing the server to perform multiplication with the encrypted $x'$.  On the other hand, if $w$ is private to the server, the server would perform the computations with the encrypted $x'$ and $w'$, and then add noise to the intermediate results $[f([x'], w)]$ through the noise flooding technique~\cite{FHE-Gentry2009} to enhance the privacy of the function.

\begin{figure}[tbp]
\centerline{
  \hspace{0.3cm}
\includegraphics[width=0.45\textwidth]{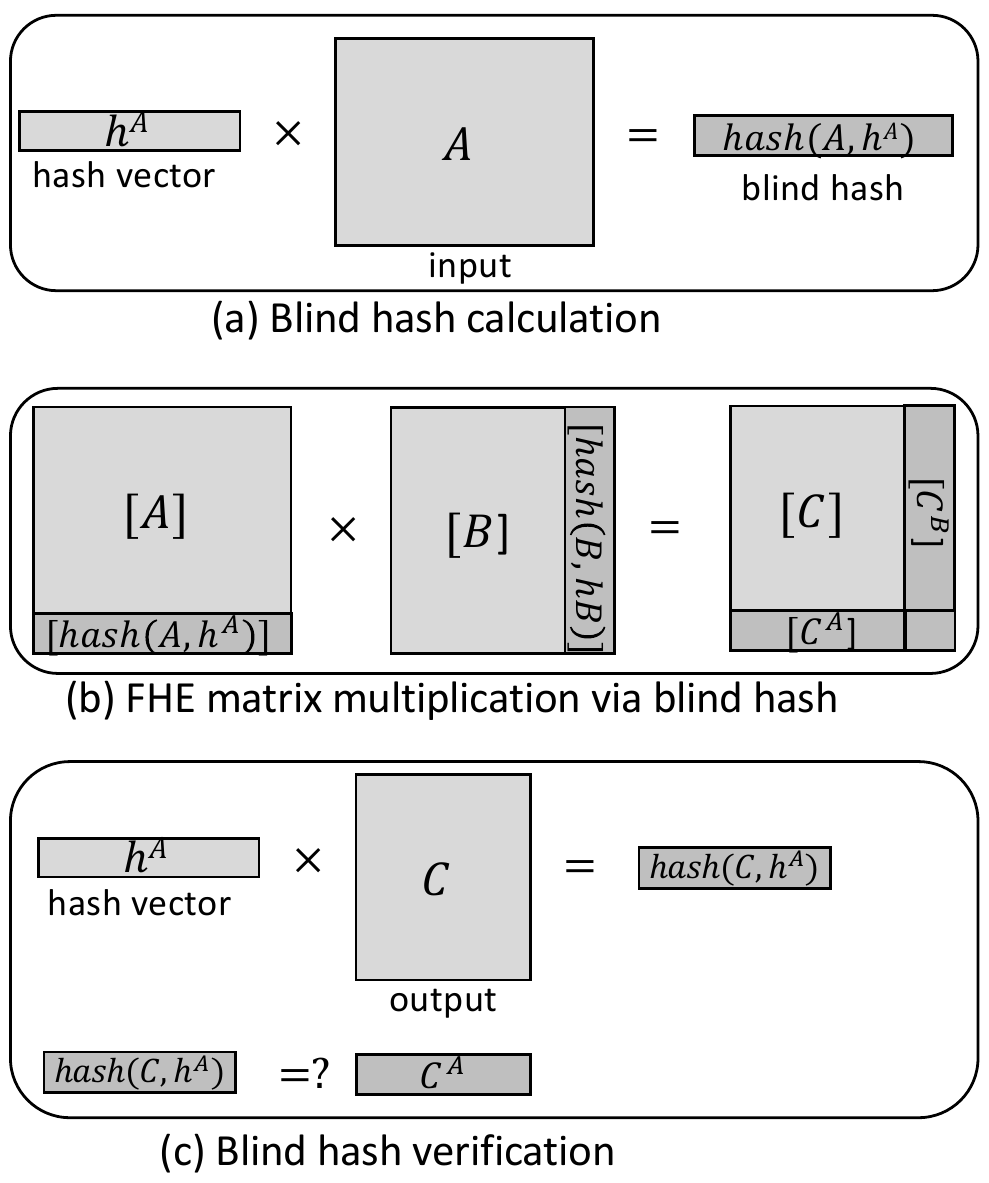}}
\caption{Illustrating vFHE with blind hash, which enables the verification of Fully Homomorphic Encryption (FHE) against both malicious tampering and computational errors. }
\label{fig:illustration}
\end{figure}

\textbf{Blind Hash Illustration.}
We illustrate the process of a blind hash by utilizing matrix multiplication, as depicted in Figure~\ref{fig:illustration}. Assume the data owner has matrix $A$ with a size of $m\times n$ and requests matrix multiplication service of $C=A \times B$  from the server, and $B$ has a length of $n\times k$. %In this example, the variable $B$ is not encrypted. However, its encrypted scenario can also be accommodated by adhering to the privacy protection steps outlined previously.  
Directly encrypting $A$ as $[A]$ using FHE and allowing the server to calculate the matrix multiplication of $[A]$ and $B$ may pose integrity risks for two reasons. Firstly, the server can manipulate the outcome $C$ into $C'$, or even fabricate a result without actually performing the computation. Secondly, the FHE computations executed by the server are susceptible to errors, including excessive FHE noise and hardware malfunctions. While ABFT~\cite{hari2021making-checksum, ABFT-1984-checksum, ding2011matrix-2011-checksum, checksum-2017} effectively detects errors, it falls short in detecting attacks in untrusted environments. {\em Blind hash} not only detects faults, as in ABFT but also detects attacks since it incorporates an extra layer of security by adding a blind hash function into the checksum computation process. 

Our blind hash method ensures Fully Homomorphic Encryption (FHE) integrity by following these steps: The client creates a hash encoding vector $h^A$ with dimensions $1\times m$, where $m$ represents the number of rows in data $A$, and each element is randomly chosen from the plaintext space. The client then multiplies $h^A$ by $A$ to produce the hashed value $hash(A, h^A)$, as depicted in Figure~\ref{fig:illustration}(a). The client encrypts the pair $(^A_{hash(A, h^A)})$ into ciphertext $[A]$ using FHE and shares this with the server. This FHE encryption guarantees that the hashed checksum value $hash(A, h^A)$ remains hidden.

The server carries out FHE matrix multiplication between $[A]$ and $B$, resulting in the encrypted outcome and proof, denoted as $([^C_{C^A}])$, where $[C]$ is the result and $C^A$ is the computational proof, as demonstrated in Figure~\ref{fig:illustration}(b).

Upon receiving the encrypted outcome and proof, the client decrypts them into plaintexts $C$ and $C^A$. The client then multiplies $C$ by the blind hash vector $h^A$ to compute $hash(C, h^A)$. The integrity of the computation is confirmed by comparing $hash(C, h^A)$ with $C^A$. If they match, the computation is deemed valid. Otherwise, an integrity issue exists. This procedure is illustrated in Figure~\ref{fig:illustration}(c).

\section{Scheme Analysis}
\textbf{Security Analysis}. If the hash vector $h^A$ is exposed to other parties, such as the server, the checksum will no longer be considered {\em blind}, and its integrity can be compromised. For instance, the server could add $h^AM$ to both $C$ and $h^AC$ simultaneously, where $M$ shares the same size as $C$. This type of manipulation would go undetected by the client. In our blind hash, however, $h^A$ is kept hidden from the server as $h^AA$ is encrypted using FHE, and the encrypted $[h^AA]$ is transmitted to the server. Without knowledge of $h^A$, it becomes difficult for the server to attack the checksum verification. 

\textbf{Ensuring Security}.
One might contend that the blind hash method is not secure if the data matrix is invertible, allowing an attacker to extract the value of $h^A$ by calculating the matrix inverse $A^{-1}$ and applying $h^AAA^{-1}$ to obtain $h^A$. We propose two solutions to address this concern. First, since only a square matrix is invertible, we alter matrix $A$ to be non-square. For example, the client can configure the FHE scheme such that the number of slots exceeds the data size of $A$. Second, the client can introduce errors into the blind hash, which serves to impede the extraction of $A$'s inverse.

\begin{figure}[tbp]
\centerline{
  \hspace{0.3cm}
\includegraphics[width=0.5\textwidth]{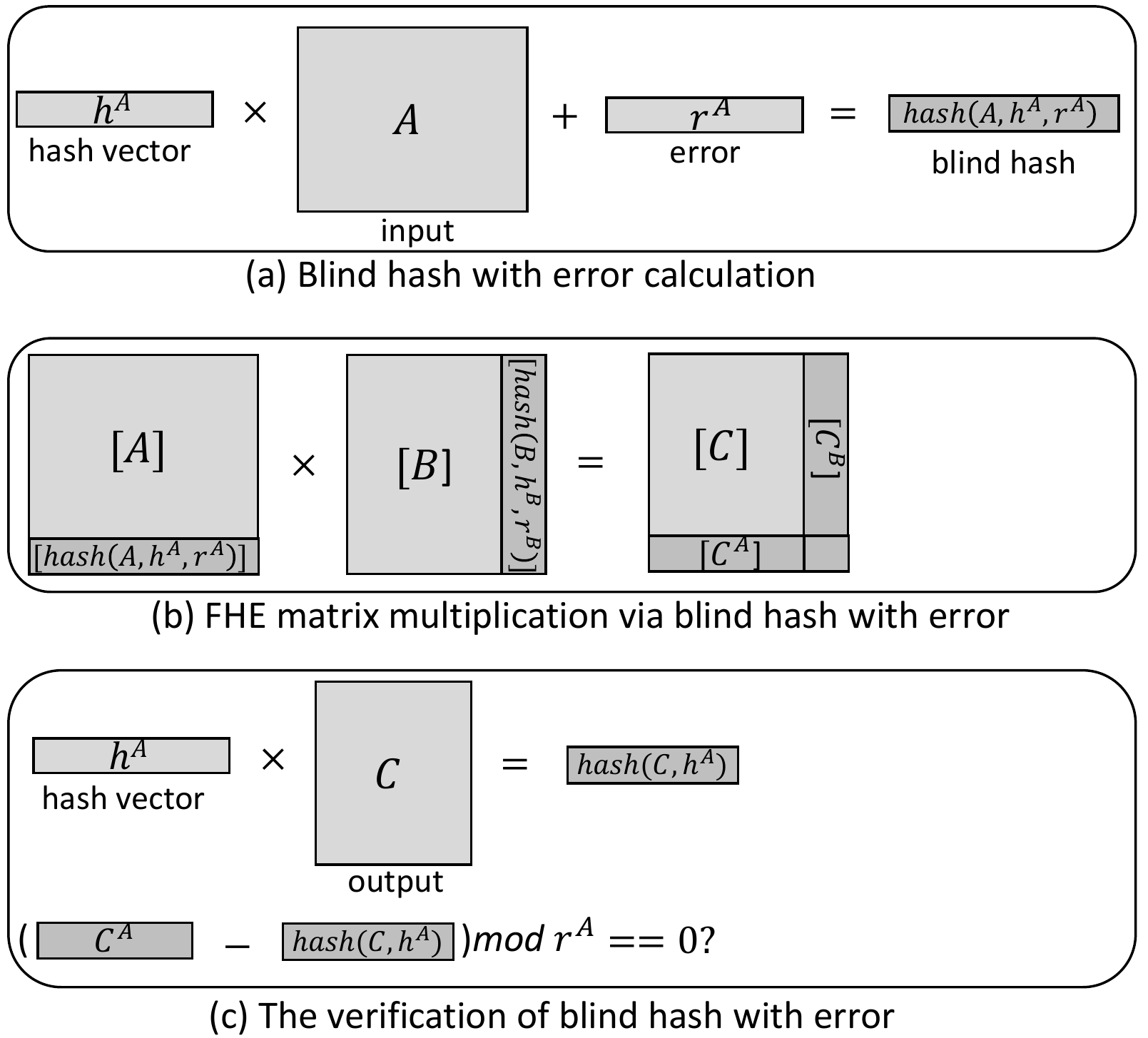}}
\caption{Blind Hash with Error enhances the security of Blind Hash}
\label{fig:extended}
\end{figure}

In the first approach, the blind hash functions effectively when the matrix is non-square. If the matrix is square, one can either partition it into two non-square matrices or introduce padding to transform it into a non-square matrix.
%If the number of slots is more significant than the number of data, there is a padding process during encryption. FHE scheme will pad many new numbers to fill all of the slots that transform the dimension of the matrix. A data matrix that was previously invertible in the plaintext space would become non-invertible in the ciphertext space. To compute the inverse matrix, an attacker must first decrypt the ciphertext data, which is impossible since the decryption process requires the secret key.
In the second approach, we introduce a blind hash with the error shown in Figure~\ref{fig:extended} based on the original blind hash. 
Given a matrix $A$, the data owner calculates the blind hash $hash(A, h^A, r^A)$ by $h^AA+r^A$ where $r^A$ is the \textit{error}, i.e., a random secret vector as the Figure~\ref{fig:extended}(a) shows.  

Subsequently, the data owner appends the $hash(A, h^A, r^A)$ to matrix $A$ and encrypts them together. The same procedure can be applied to matrix $B$, as illustrated in Figure~\ref{fig:extended}(b). The FHE computation step in this case is identical to that of the blind hash. During verification, the data owner must execute two steps. First, the data owner computes the $hash(C, h^A)$, resulting in $h^AC$. Second, the data owner subtracts $hash(C, h^A)$ from $C^A$ and modulo the error $r^A$, and verifies if the outcome equals 0, as shown in Figure~\ref{fig:extended}(c).

%Then the client can verify the integrity by subtracting $e^TAB$ and doing a modulus operation with secret vector $r^T$, as shown in figure \ref{fig:extended} (b). The secret vector $r^T$ prevents the attacker from extracting the hash key $e^T$ even though they have the invertible matrix $A^{-1}$. 

\begin{figure*}[t]
    \centering
    \includegraphics[width=0.9\textwidth]{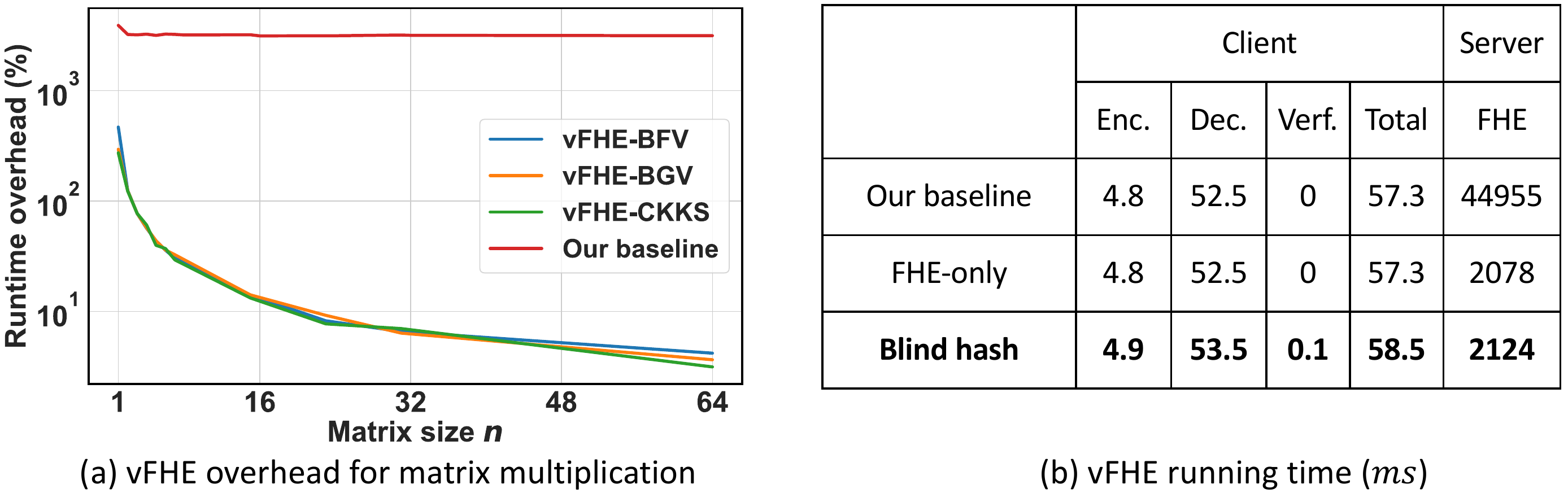}
    \caption{vFHE runtime overhead.}
    \label{fig:results}
\end{figure*}

\textbf{Overhead Analysis.} We demonstrate that our implementation of a {\em blind hash} results in minimal increases in various areas, including plaintext expansion, server-side computational overhead, ciphertext expansion, and client computation. (i) Plaintext expansion.  When applied to a matrix with dimensions of $m\times n$, the resulting {\em blind hash} will have dimensions of $1\times n$. The plaintext expansion rate of the {\em blind hash} can be calculated as $\frac{1}{m}$, indicating a relatively small increase in the size of the original matrix. (ii) FHE computation. The implementation of a blind hash does not require any modifications to the server. A {\em blind hash} has a minimal impact on computational resources, particularly in the case of large matrix multiplications. The computational complexity of multiplying two matrices, $A$ and $B$, with dimensions of $m\times n$ and $n\times k$ respectively, is expressed as $\mathcal{O}(mnk)$. A {\em blind hash} increases the computational number from $mnk$ to $(m+1)nk$, yet the overall computational complexity remains unchanged at $\mathcal{O}(mnk)$. This demonstrates that the FHE computation overhead of a {\em blind hash} is negligible for sufficiently-sized matrices. (iii) Ciphertext expansion. The ciphertext expansion rate, which determines the communicational overhead between the client and the server, is equal to or less than $1+\frac{1}{m}$ due to the ability to pack multiple values into a single ciphertext. 
(iv) Client computation. The client's computational overhead is comprised of two components: the generation of the hash vector and checksum vector, and the verification of the proof. The generation of the hash vector can be performed in advance, while the generation of the checksum vector involves a vector-matrix multiplication with a computational complexity of $\mathcal{O}(mn)$, where the vector and matrix have dimensions of $1\times m$ and $m\times n$, respectively. The overhead ratio of this process to the original computation is $\mathcal{O}({\frac{1}{k}})=\frac{\mathcal{O}(mn)}{\mathcal{O}(mnk)}$. The verification process of the proof involves a vector-matrix multiplication with a computational complexity of $\mathcal{O}(mk)$, i.e., $\frac{1}{n}$ of matrix multiplication, and a comparison of two vectors with dimensions of $1\times k$, which has a complexity of $\mathcal{O}(k)$, i.e., $\frac{1}{mn}$ of matrix multiplication. We propose the use of power-of-2 hash values and inexpensive shift operations to reduce the costly $\mathcal{O}(mn)$ multiplication with inexpensive shift operations, leading to a significant decrease in the client's computational overhead. As convolution operations in deep neural networks can be represented mathematically as matrix multiplications~\cite{checksum-2017}, we only analyze the complexity of representative matrix multiplications.

\section{Result}
\label{sec:results}

To examine the potential runtime overhead of vFHE, we have implemented vFHE of code based on SEAL~\cite{SEAL} to insert an FHE integrity verification. Previous research has demonstrated that the FHE-in-TEE approach yields exceptional efficiency for variable FHE, as reported in~\cite{viand2023verifiable}. Accordingly, our baseline is established by incorporating SEAL-based BFV and AMD-SEV TEE. Figure~\ref{fig:results}(a) shows execution time overheads as a function of the matrix size, over unprotected execution (FHE-only), with various arithmetic FHE schemes, including BFV, BGV, and CKKS. As shown in the figure, the overhead of our baseline FHE-in-TEE over the FHE-only method executing matrix multiplication is far more than $1000\%$. With {\em blind hash} support, the overhead decreases to about 4\% for diagonal matrix multiplication with size $n=64$. The runtime overhead of {\em blind hash} is reduced as the matrix size, $n$, increases, with a worst-case overhead of fewer than 5$\times$ when $n=1$. When the matrix size is large enough, its overhead will be near zero. We use Table (b) in Figure~\ref{fig:results} to show execution time comparisons of our {\em blind hash} method in vFHE with the BGV scheme against our baseline and FHE-only method on a diagonal matrix-matrix multiplication with size $n=64$. Our baseline FHE-in-TEE offers verifiable FHE without adding any additional runtime overhead on the client side about encryption (Enc.), decryption (Dec.), and verification (Verf.). However, it does substantially increase overhead on the server side, by approximately 21.6 times, compared to the FHE-only method. In contrast, the FHE computation equipped with our blind hash in vFHE increases $2.1\%$ and $2.2\%$ runtime overhead on the client and server sides, respectively, over the FHE-only method. The preliminary results demonstrate that, with the proposed techniques, vFHE with {\em blind hash} can be applied universally to various FHE schemes (BFV, BGV, CKKS) and can achieve efficiency, indicating the promise of vFHE as a new paradigm for verifiable FHE computation.
%The runtime overhead of SBCR is decreased as the matrix size $n$ and its worst-case overhead when $n=1$ is still less than $5\times$. The preliminary results confirm that with the proposed techniques, SBCR can be universal to various FHE schemes and can be made efficient, indicating the promise of SBCR as a new paradigm for verifiable FHE computation.  

%region-based analysis on LLVM to insert attach and detach into a given C program. Based on Sniper~\cite{carlson2011etloafsaapms}, we built a prototype of the hardware support of the fast attach/detach mechanism~\cite{Xu+:ASPLOS20}, and also the scheme to silent unnecessary detach calls as described in Section~\ref{sec:architecture}. The hardware simulator is presented in Table (a) in Figure~\ref{fig:results}. WHISPER benchmarks~\cite{nalli2017analysis} are used.
%Figure~\ref{fig:results}(b) shows execution time overheads over unprotected execution, broken down into components from attach and detach system calls, re-randomization (Rand), execution of conditional attach/detach instructions (Cond) and other overheads (e.g., permission matrix). As  shown in the figure, the overhead of executing every attach/detach with system calls averages 50\%. With TERP architecture support, the overhead decreases to about 6\% with 40$\mu$s exposure window (EW) and as low as 2\% with 160$\mu$s EW. The preliminary results confirm that with the proposed techniques, TERP controls can be made efficient, indicating the promise of TERP as a new paradigm for memory protection.  

\section{Conclusion}
In this research paper, we present the blind hash, an innovative approach designed to ensure data integrity in FHE computations. Our proposed method offers several key advantages, including scalability, low computational overhead, and compatibility with a diverse range of popular FHE schemes currently available. By addressing data integrity concerns, the blind hash technique contributes to the growing body of research on secure FHE computations and has the potential to enhance the practicality of FHE-based solutions across numerous applications.

\bibliographystyle{plain}
\bibliography{lou, lou-neurips, lou-iclr}

\end{document}